\documentclass{article}
\usepackage[utf8]{inputenc}
\usepackage{amsmath}
\usepackage{amssymb}
\usepackage{graphicx}
\usepackage{lscape}
\usepackage{geometry}
\usepackage{xcolor}
\usepackage{authblk}

\title{Reproducing size distributions of swarms of barchan dunes on Mars and Earth using a mean-field model}
\author[1]{Dominic T Robson}
\affil[1]{\textit{{\footnotesize King's College London, Department of Geography, Bush House, North East Wing, 40 Aldwych, London, UK, WC2B 4BG}}}

\author[2]{Alessia Annibale}
\affil[2]{\textit{{\footnotesize King's College London, Deparment of Mathematics, Strand Building, Strand Campus, Strand, London, UK, WC26 2LS}}}
\author[1]{Andreas CW Baas}

\date{August 2022}

\begin{document}

\maketitle

\begin{abstract}
    We apply a mean-field model of interactions between migrating barchan dunes, the CAFE model, which includes spontaneous calving, aggregation, fragmentation, and mass-exchange, yielding a steady-state size distribution that can be resolved for different choices of interaction parameters. The CAFE model is applied to empirically measured distributions of dune sizes in two barchan swarms in the north circumpolar region of Mars, three swarms in Morocco, and one in Mauritania, each containing more than 1000 bedforms.  When the sizes of bedforms are rescaled by the mean size in each zone two attractor states appear, with the Tarfaya zones all displaying a common distribution and the Martian and Mauritanian zones sharing a different distribution.  Comparison of these attractor states with the outputs of the CAFE model reveals that the Tarfaya-type distribution results from a preference for aggregation and fragmentation interactions whereas the Mars-Mauritania distribution is more likely a result of exchange-dominated interactions.  We observe that there appears to be a greater number of collisions happening in Tarfaya than in the other areas which is consistent with a greater rate of aggregation-fragmentation processes as suggested by our model.  Our comparison with the CAFE model also predicts a universal rule for the outputs of the different types of interactions with exchange interactions favouring the production of two dunes roughly equal in size.  Fragmentation interactions often result in three bedforms with volumes approximately in the ratio 1:1:2.  Finally, we find that spontaneous calving of dunes does not play an important role in shaping the size distributions in barchan swarms.
\end{abstract}

\maketitle

\section{Introduction}

Barchan dunes are a class of migrating aeolian bedform that are found in many areas on Mars and Earth under unidirectional wind regimes and relatively low sediment supply conditions \cite{bourke2009varieties, goudie2020global}.  Under such a near-constant wind direction, the dunes form into their striking crescentic shape with steep slip-faces resulting from avalanching \cite{bagnold1954}.  Properties of individual barchans, such as relationships between migration rate and size or between the various morphometric parameters, are well-known and can be thought of as emergent properties of aeolian sand transport \cite{bagnold1954, elbelrhiti2008barchan, gay1999observations, seif2019desertification, hamdan2016morphologic, amin2019environmental, yang2020experimental, yang2019migration, hu2019origin}.\\

Simulations have shown, however, that the size of isolated barchans is not stable and that the bedform should either grow indefinitely or shrink and vanish due to inevitable imbalance between incoming sand flux (gain) at the upwind toe of the dune and outgoing sand flux (loss) from its horns\cite{duran2010continuous, andreotti2002selection, hersen2004corridors, kroy2002minimal, zhang2010morphodynamics}.  In reality, barchans are not found as isolated dunes but in vast collections, known as swarms, which can contain tens of thousands of migrating dunes \cite{hersen2004corridors, elbelrhiti2008barchan, duran2009dune, duran2011size}.  Within these swarms the size distribution of barchans appears to be homogeneous in the direction of the wind \cite{elbelrhiti2008barchan}.  This homogeneity contrasts with the instability observed in simulations from which one would expect that the average size should increase linearly with downwind distance \cite{duran2011size}.  Two types of process have been proposed to explain the emergence of apparent stability of dune sizes in barchan swarms: calving, where a small dune spontaneously sheds off of the horn of a larger one \cite{elbelrhiti2008barchan, elbelrhiti2012initiation, worman2013modeling}, and dune-dune interactions such as exchange collisions \cite{duran2009dune, duran2011size, assis2020comprehensive, diniega2010long, genois2013spatial}.\\

There is significant variation in the terminology of barchan interactions.  In our work calving is defined as a purely spontaneous $1 \rightarrow 2$ process, whereas the term has also sometimes been applied to the process whereby fragmentation of a larger barchan is induced by an incoming upwind dune \cite{elbelrhiti2012initiation}.  Induced fragmentation of a dune is a $2 \rightarrow 3$ process and in our study we refer to this as an \textit{fragmentation interaction}.  We apply the same term for any $2 \rightarrow 3$ collision, covering various terminology from previous studies such as: ``budding'', ``breeding'', and ``splitting'' to name but a few \cite{duran2009dune, assis2020comprehensive, katsuki2011cellular, endo2004observation}.  We discuss the disparate vocabulary in the field and the terminology of this study further in section 2. \\

Barchan swarms sit at the top of a hierarchical structure of emergent phenomena.  Migration rates and morphological characteristics of barchans emerge from the mechanisms of aeolian sand transport.  In turn,  properties of swarms such as peaked size distributions and the alignment of dunes with the horns of their upwind neighbour are emergent phenomena associated with the interactions between the bedforms.  Owing to the long timescales over which these processes occur in real-world dune fields, studies of these interactions are typically performed using computer simulations \cite{duran2009dune, duran2010continuous, diniega2010long, katsuki2011cellular} or in water-tank studies \cite{assis2020comprehensive, bacik2020wake, endo2004observation}.  Although these techniques allow the investigation of the entire processes of dune-dune interactions, they cannot be used to study systems of more than $\sim 10$ bedforms at a time \cite{duran2011size}.\\

In order to study larger populations, one must instead turn to mean-field \cite{duran2009dune, genois2013spatial} or agent-based models \cite{duran2011size, worman2013modeling, genois2013spatial, lima2002modelling, parteli2003simple, lee2005modelling, diniega2010long}.  Collectively we will refer to these as \textit{swarm-scale} models.  Broadly speaking, swarm-scale models have been successful in replicating certain properties of real-world populations of dunes, but there are significant limitations in all of the existing models.  Mean-field models are simpler than agent-based models in that they neglect any spatial dependence.  The most widely cited mean-field model of barchan swarms found that a collision rule for exchange interactions derived from continuum simulations leads to an approximately Gaussian distribution of barchan widths \cite{duran2009dune}.  A simplistic description of sand-flux was then included to alter the dune sizes in the model in order to replicate the distributions observed in several real-world swarms \cite{duran2009dune}.  Despite this apparent success, when the same interaction rules were implemented in an agent-based model, the resulting size distribution was found to be heterogeneous \cite{duran2011size}.  In fact, the average size of dunes in the agent-based model increases linearly with downwind distance \cite{duran2011size} as would be the case even if collisions were not included.  Several other agent-based barchan swarm models \cite{worman2013modeling, lima2002modelling} have also failed to replicate the downwind homogeneity of real-world dune fields \cite{elbelrhiti2008barchan}.  Some models produce size distributions which display features that are not observed in real-world swarms.  For example, a significant peak at a single large dune size defined by a calving threshold rule \cite{worman2013modeling} or absence of any dunes larger than the size at which they are introduced \cite{genois2013spatial}.  The latter directly contradicts our understanding that new barchans initiate in a swarm at the minimum dune size \cite{elbelrhiti2012initiation}, and therefore these incipient dunes are the smallest rather than the largest in a swarm.  The mixed success of barchan swarm-scale models is likely due to the fact that each of the models described above include some but not all of the known dune-dune interactions.\\ 

In this work we explore a mean-field model which allows for calving, as well as three different types of barchan-barchan interactions which are known to occur in these swarms.  The model is built from a general mean-field model which we developed recently and which can be applied to understand the steady-state distributions of collections of interacting bodies \cite{robson2021combined}.  The model unites continuous aggregation-fragmentation \cite{krapivsky2010kinetic, fowler2016theoretical} and asset-exchange models \cite{lux2005emergent} and, as such, can include any $n \rightarrow m$ process in which an extrinsic quantity is conserved.  Despite the generality of the processes permitted in the model, it is possible to analytically derive the integer moments of the steady-state distribution of the conserved quantity as well as find a self-consistency equation for the steady-state distribution itself.  Although any processes can be included, for this work we limited ourselves to only those processes which have been observed in experiments and simulations.  Our model therefore represents an extension to the existing mean-field models \cite{duran2009dune, genois2013spatial} as neither of the previous models has included calving.  Furthermore, we compare the steady-state outputs of this modelling to real-world size distributions which we measured for several locations on both Mars and Earth, whereas comparison of real-world swarms has only been conducted previously using a mean-field model that omitted fragmentation and focused predominantly on exchange \cite{duran2009dune}.  Additionally, because the link between outputs and inputs of our model are known we can gain physical insight into the nature of the processes in barchan swarms beyond what was possible in previous mean-field modelling.\\

In section \ref{sec: Model} we provide a description of the mean-field model and tailor the results of the generalised model \cite{robson2021combined} to the processes which are relevant to barchan swarms.  We then describe, in section \ref{sec: Measurements}, the methods by which we measured the sizes of the real-world barchans and the locations of the swarms we studied.  In section \ref{sec: Results} we show the distributions we observed for the swarms, describe how we then optimised the model parameters to reproduce these, and show the results of the optimisation, reserving discussion of the physical relevance of these findings and how the work may be improved to sections \ref{sec: Discuss} and \ref{sec: Conclusion}.

\section{Mean-field modelling}\label{sec: Model}

  Mean-field models are useful tools in studying the global properties of large populations of interacting elements.  The model we apply in this work is a specific implementation of the general mean-field model which we have described in \cite{robson2021combined}.\\

\subsection{Interactions and calving}

We consider a system comprising $N(t)$ interacting particles (dunes) at time $t$.  The dunes are characterised by their volume, a continuous quantity distributed according to a volume probability density function (pdf) $p(v,t)$.  Note here that, since the bulk density will be approximately the same in all the barchans within a field, volume is equivalent to mass.\\

\begin{figure}
    \centering
    \includegraphics[width = \textwidth]{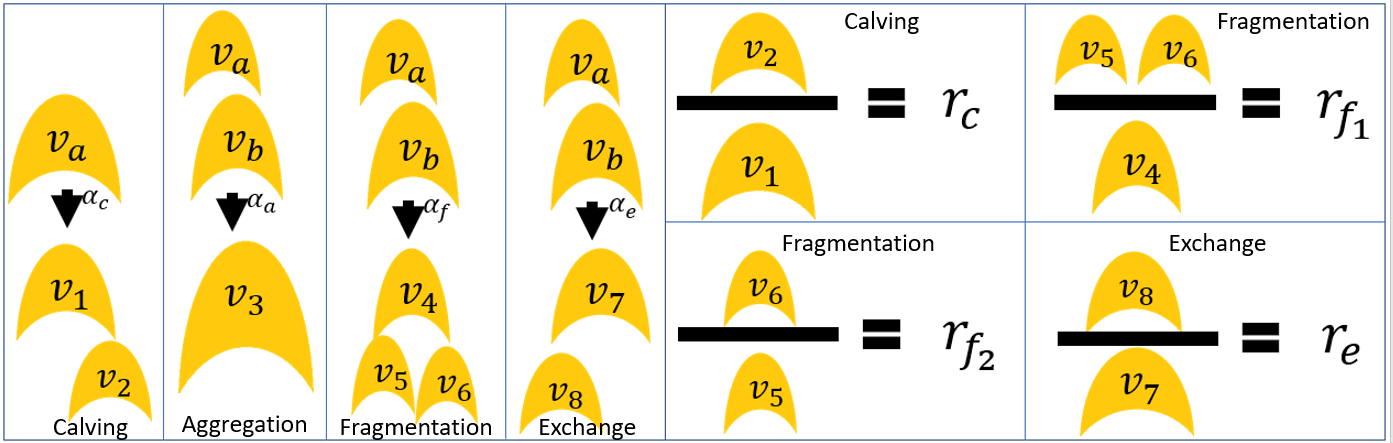}
    \caption{The four processes (left to right: calving, aggregation, fragmentation, and exchange), eight output channels of the  {CAFE} model ($v_{1,...,8}$), and the definitions of the stochastic variables ($r_c$, $r_{f_1}$, $r_{f_2}$, and $r_e$) which define the interaction rules.  The rate coefficients $\alpha_c$, $\alpha_a$, $\alpha_f$, $\alpha_e$ of the four processes are also shown.}
    \label{fig: Processes}
\end{figure}

In the general model \cite{robson2021combined} we allowed for any $n \rightarrow m$ processes in which volume is conserved.  Here we limit ourselves to only those processes which are known to be relevant in barchan swarms, focusing on four interactions which are represented pictorially in figure \ref{fig: Processes}.  The four processes can be described as follows:     

\begin{itemize}
    \item \textit{Calving} - a $1 \rightarrow 2$ process where a dune spontaneously sheds a portion of it volume as a new dune.
    \item \textit{Aggregation} - a $2 \rightarrow 1$ process where two dunes merge together.
    \item \textit{Fragmentation} - a $2 \rightarrow 3$ process where two dunes break into three.
    \item \textit{Exchange} - a $2 \rightarrow 2$ process where volume is transferred between the dunes.
\end{itemize}

Calving is thought to occur in barchan swarms due to changes in wind regime or storm events which can destabilise larger dunes resulting in a smaller barchan breaking off of the flank \cite{elbelrhiti2008barchan, elbelrhiti2005field}.  This process is shown in the first panel of figure \ref{fig: Processes}.  Barchans have been observed to aggregate, usually referred to in dune literature as ``merging'' or ``coalescence'' ,in real-world fields \cite{gay1999observations} as well as in simulations \cite{duran2009dune, duran2010continuous, duran2011size} and water-tank experiments \cite{assis2020comprehensive}.  We show how we treat the aggregation process in the second panel of figure \ref{fig: Processes}.\\

Fragmentation interactions have also been observed in various settings \cite{ duran2009dune, duran2010continuous, duran2011size, assis2020comprehensive, elbelrhiti2012initiation} with several different types of fragmentation having been identified for example``fragmentation-chasing'' and ``fragmentation-exchange'' \cite{assis2020comprehensive}, ``budding'' and ``breeding'' \cite{duran2005breeding, duran2009dune, duran2010continuous}, and ``splitting'' \cite{endo2004observation, katsuki2011cellular}.  All of these processes are $2 \rightarrow 3$ interactions which we refer to as fragmentation.  The fragmentation process we consider in the our modelling is shown in figure \ref{fig: Processes}. A few studies use the term ``calving'' in reference to airflow turbulence in the wake of a nearby dune driving a barchan to split \cite{elbelrhiti2008barchan, elbelrhiti2012initiation}.  Although this dune-induced process has been referred to as ``calving'', the requirement of a nearby upwind dune means that this is a $2 \rightarrow 3$ process and therefore will be include in our fragmentation interactions and not in the calving process.  We will use the term ``calving'' to represent only the spontaneous breaking of dunes, a $1 \rightarrow 2$ process as shown in the first panel of figure \ref{fig: Processes}.\\

Finally, the exchange interaction has also been observed in nature \cite{vermeesch2011solitary}, simulations \cite{duran2009dune, duran2010continuous, duran2011size, schwammle2003solitary, endo2004observation, hugenholtz2012real, katsuki2011cellular}, and experiments \cite{assis2020comprehensive}.  While commonly referred to as exchange \cite{assis2020comprehensive}, $2 \rightarrow 2$ interactions have also been variously termed ``solitary-wave behaviour'' \cite{schwammle2003solitary, vermeesch2011solitary}, ``ejection'' \cite{endo2004observation, hugenholtz2012real, katsuki2011cellular}, and ``reorganization'' \cite{katsuki2011cellular}.  We use the term exchange to represent all $2 \rightarrow 2$ processes.  The exchange process is diagramatised in the fourth panel of \ref{fig: Processes}.\\

Since the model we are describing in this work features only these four processes we will refer to it as the  {C}alving,  {A}ggregation,  {F}ragmentation, and  {E}xchange ({CAFE}) model.  All four of these processes have been found to play a role in the formation of non-trivial size distributions in real-world systems \cite{krapivsky2010kinetic, fowler2016theoretical, lux2005emergent}\\

For the  {CAFE} model we assume that all four processes conserve volume so that the total volume of all the dunes in the system is conserved.  In real-world systems, we believe that these processes ought to be at least approximately conservative, and this assumption is in line with previous agent-based and mean-field modelling of barchan fields \cite{duran2009dune, duran2011size, genois2013spatial, worman2013modeling, diniega2010long, lima2002modelling}.\\

\subsection{Output channels and interaction rules}

In \cite{robson2021combined} we introduce the concept of \textit{output channels} (OCs) which is pivotal in our derivations.  OCs give expressions, based on the choice of rules for the individual outputs of all the four processes.  There are two outputs of calving, one of aggregation, three of fragmentation, and two of exchange so there are eight OCs in the {CAFE} model.\\

Consider a dune of volume $v_a$ calving to yield dunes with volumes $v_1$ and $v_2$ these are OC1 and OC2.  Volume conservation tells us that $v_1 + v_2 = v_a$ but we need one additional piece of information to determine the two volumes.  This additional piece of information is the rule for the calving process, in this case, the ratio between the two output dunes, $r_c$.  To derive analytical results only the rules must be random i.e. independent of the input volume \cite{robson2021combined}.  The rule for calving is the probability distribution, $p_c(r_c)$, from which the stochastic variable $r_c = v_2/v_1$ is drawn.  We now have enough information to define the two OCs associated with calving

\begin{equation}
    v_1 = \frac{v_a}{1 + r_c} \quad \text{and} \quad v_2 = \frac{v_a r_c}{1 + r_c}.
\end{equation}

\noindent{For the remainder of the {CAFE} model processes, we consider inputs $v_a$ and $v_b$ and introducing additional stochastic variables $r_{f_1}$, $r_{f_2}$, and $r_e$ drawn from distributions $p_{f_1}(r_{f_1})$, $p_{f_2}(r_{f_2})$, and $p_{e}(r_{e})$.  The OCs are then defined as for aggregation,}

\begin{equation}
    v_3 = v_a + v_b,
\end{equation}

\noindent{for fragmentation,} 
\begin{align}
        v_4 &= \frac{v_a + v_b}{1 + r_{f_1}},\\
        v_5 &= \frac{(v_a + v_b)r_{f_1}}{(1 + r_{f_1})(1 + r_{f_2})},\\
        v_6 &= \frac{(v_a + v_b)r_{f_1}r_{f_2}}{(1 + r_{f_1})(1 + r_{f_2})},
\end{align}

\noindent{and for exchange collisions}

\begin{align}
        v_7 &= \frac{v_a + v_b}{1 + r_e},\\
        v_8 &= \frac{(v_a + v_b)r_e}{1 + r_e}.
\end{align}

\subsection{Rates and channel probabilities}\label{sec: Rates}

Evaluating the time evolution of $p(v,t)$ requires understanding how often the OCs create a dune of volume $v$.  We have already defined the OCs but have yet to discuss the rate at which the interactions occur.\\

Since we are considering a mean-field model, in any $n$-body process, all combinations of $n$-dunes must be equally likely to be involved.  At time $t$ there are $\binom{N(t)}{n}$ such combinations and therefore the rate at which $n$-body processes occur is proportional to $\binom{N(t)}{n}$.  We refer to the proportionality constants as \textit{rate coefficients} which we label $\alpha_c$, $\alpha_a$, $\alpha_f$, and $\alpha_e$ for calving, aggregation, fragmentation, and exchange respectively.  OC1 and OC2 are outputs of calving, a one-body process, while the other OCs are results of two-body processes i.e. barchan-barchan interactions.  We can now write the channel probabilities $p_i$ which are given as the rate of a particular channel divided by the total rates of all channels $\alpha_{out}$ \cite{robson2021combined}

\begin{align}
    p_1 &= p_2 = \frac{\alpha_c}{\alpha_{out}}N(t),\\
    p_3 &= \frac{\alpha_a}{2\alpha_{out}}N(t)(N(t) - 1),\\
    p_4 &= p_5 = p_6 = \frac{\alpha_f}{2\alpha_{out}}N(t)(N(t) - 1),\\
    p_7 &= p_8 = \frac{\alpha_e }{2\alpha_{out}}N(t)(N(t) - 1),
\end{align}

\noindent{where}

\begin{equation}
    \alpha_{out} = 2\alpha_c N(t) + \frac{\alpha_a + 3\alpha_f + 2\alpha_e}{2}N(t)(N(t) - 1).
\end{equation}

\noindent{Channel probabilities represent the fractions of dunes in the population that were generated through each OC, while $\alpha_{out}$ gives the total rate at which outputs are generated.  Together with the definitions of the output channels, we have now completely defined the {CAFE} model.  In the four leftmost panels of figure \ref{fig: Processes} we show, diagrammatically, the four processes and how they relate to the OCs and rate coefficients.  The diagrams to the right of figure \ref{fig: Processes} show the definitions of the stochastic variables in terms of the OCs.}\\

The free parameters of the model are the rate coefficients, distributions of the stochastic variables, and the total volume of all the dunes.  If we know the steady-state population size, $N_s$, not all of the rate coefficients are free parameters since the steady population size requires a balance between the additive processes (fragmentation and calving) and the reductive ones (aggregation).  In the {CAFE} model, the steady-state population size is

\begin{equation}
    N_s = 1 + \frac{2\alpha_c}{\alpha_a - \alpha_f},
\end{equation}

\noindent{which we use as a constraint on $\alpha_a$.}

\subsection{Master equation and steady-state probability density function}

Having completely defined the model we can now insert all of the terms into the general equations we derived in \cite{robson2021combined}.  The evolution of $p(v,t)$ is determined by the balance between the creation of dunes of volume $v$ through the output channels, and the loss of dunes of volume $v$ due to such dunes being involved in interactions.  The balance of the loss and gain terms combine to give a master equation for the time-derivative of $p(v,t)$

\begin{equation}\label{eq: MasterEquation}
    \dot{p}(v,t) = \frac{\alpha_{out}}{N(t)}\big(p_{gain}(v,t) - p(v,t) \big),
\end{equation}

\noindent{where}

\begin{equation}
    p_{gain}(v,t) = \big\langle \delta\big(v_i(v_a, v_b, r_c, r_{f_1}, r_{f_2}, r_e) - v\big) \big\rangle_{i, v_a, v_b, r_c, r_{f_1}, r_{f_2}, r_e},
\end{equation}

\noindent{with $\delta(x)$ denoting the Dirac $\delta$-function and $\langle \cdot \rangle_x$ the average over the distribution of $x$.  Averaging over the OCs is done using the channel probabilities, $p_i$.  Note that $p_{gain}(v,t)$ involves averaging over the distributions of $v_a$ and $v_b$ which are the volume pdf $p(v,t)$ if we assume that the population is large enough such that their distributions can be thought of as independent.}\\

It is trivial to write an expression for the steady-state volume pdf $p_s(v)$ by setting $\dot{p}(v,t) = 0$ in equation \eqref{eq: MasterEquation} which gives 

\begin{equation}\label{eq: Steady}
    p_s(v) = \big\langle \delta\big(v_i(v_a, v_b, r_c, r_{f_1}, r_{f_2}, r_e) - v\big) \big\rangle^{(s)}_{i, v_a, v_b, r_c, r_{f_1}, r_{f_2}, r_e},
\end{equation}

\noindent{where we have used the superscript $(s)$ to denote that the distributions of $v_a$ and $v_b$ on the right hand side of \eqref{eq: Steady} are also the steady-state distribution $p_s(v)$.  This is a self-consistency equation as the steady-state pdf appears on both sides.  The form of the self-consistency equation means, however, that it can easily be solved using a population dynamics algorithm \cite{robson2021combined, mezard2001bethe, agliari2013immune}.}

\subsection{Steady-state moments}

As well as a self-consistency equation for the steady-state pdf, it is also possible to write exact expressions for the steady-state moments \cite{robson2021combined}.  As we have already stated, the total volume of dunes in the system is a free parameter, which, when divided by the steady-state population size $N_s$, gives the mean volume $\langle v \rangle_s$, where the subscript, $s$, denotes that this is in the steady-state.  The higher integer moments $\langle v^{\ell} \rangle_s$ for $\ell \geq 2$ can then be calculated iteratively \cite{robson2021combined} using

\begin{equation}\label{eq: Moments}
    \langle v^\ell \rangle_s = \frac{K_\ell}{Z_\ell} \sum_{j = 1}^{\ell - 1}\binom{\ell}{j}\langle v^j \rangle_s \langle v^{\ell - j} \rangle_s,
\end{equation}

\noindent{where}

\begin{equation}\label{eq: KL}
    K_\ell = \frac{N_s (N_s - 1)}{2\alpha_{out}^{(s)}}\left(\alpha_a + \alpha_f \left\langle \frac{1 + r_{f_1}^\ell\frac{1 + r_{f_2}^\ell}{(1 + r_{f_2})^\ell}}{(1 + r_{f_1})^\ell} \right\rangle_{r_{f_1}, r_{f_2}} + \alpha_e \left\langle \frac{1 + r_e^\ell}{(1 + r_e)^\ell}\right\rangle_{r_e} \right)
\end{equation}

\noindent{and where} 

\begin{equation}
    Z_\ell = 1 - 2K_\ell - \frac{\alpha_e N_s}{\alpha_{out}^{(s)}}\left\langle \frac{1 + r_c^\ell}{(1 + r_c)^\ell}\right\rangle_{r_c}.
\end{equation}

\section{Measuring size distributions of real-world barchan swarms}\label{sec: Measurements}

In this section, we describe the locations of the real-world barchan swarms which we have measured and the method we used for extracting the sizes of the dunes.  We first explain how the areas of study were chosen before moving on to explain how the volumes were calculated for both the barchan and non-barchan bedforms within these areas.  This includes the derivation of a scaling-law for barchan length and basal area.

\subsection{Zones of study}

A swarm of barchan dunes is a two-dimensional many-body system in which all bedforms migrate in approximately the same direction, that of the dominant wind.  Typically, mean-field models such as ours cannot be applied to such low-dimensional systems because they rely upon an assumption that the systems are well-mixed \cite{krapivsky2010kinetic}.  However, large swarms of barchans are thought to be homogeneous in the direction of the wind \cite{elbelrhiti2008barchan, duran2011size}.  As such, one can effectively treat a zone within the bulk of a large swarm as having periodic boundaries, so that the assumption of sufficient mixing is valid within the zone.  It has already been shown that the global size distributions of agent-based models of barchans can be predicted from mean-field models \cite{duran2009dune, duran2011size, genois2013spatial} which further suggests that mean-field models can be applied despite the low-dimensionality of barchan swarms.\\

\begin{figure}
    \centering
    \includegraphics[width = \textwidth]{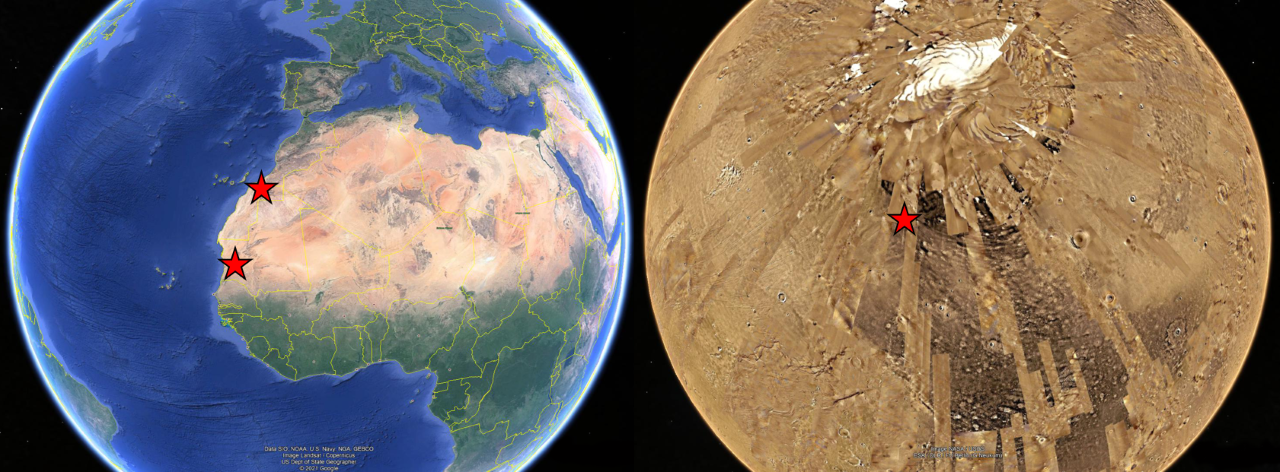}
    \caption{The locations of the zones of study on Earth and on Mars are marked as stars.  On Earth, the more northerly star covers the locations of the three zones in Tarfaya Province while the more southerly marker covers the field in Mauritania.  Image courtesy of Google Earth.}
    \label{fig: Locations}
\end{figure}

Although barchan swarms occur at many locations on Earth \cite{goudie2020global} and Mars \cite{bourke2009varieties}, in most of these locations the populations of barchans are limited in number and typically situated between different classes of bedforms such as barchanoid ridges, meaning that the assumption of periodic boundaries may not hold.  We limit ourselves to swarms in which we could define zones of $\sim1000$ barchans such that the dunes immediately upwind of the zone appeared qualitatively similar to those within the zone.  For this study, we selected six zones: three located approximately 17km west of El Hagounia, Tarfaya Province in a large field which extends southward from the northern Atlantic coast of Morocco; one around 60km south of Akjoujt, Mauritania in a field that extends southwest towards the coast; and two oriented east-southeast in the northern circumpolar region of Mars.  The locations of these six zones are shown in figure \ref{fig: Locations}.  Satellite imagery was analysed in Google Earth pro and derive from 2022 CNES Airbu and 2022 Maxar Technologies for the Earth imagery of Tarfaya and Mauritania zones, and from the 2011 Mars Reconnaissance Orbite (MRO) Context Camera (CTX) for the Mars imagery.\\

\begin{figure}
    \centering
    \includegraphics[width = \textwidth]{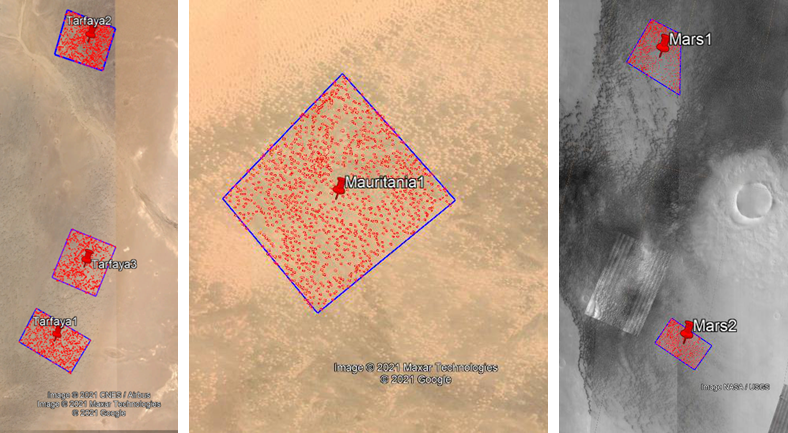}
    \caption{The size zones of study are marked as blue quadrilaterals with the bedforms inside marked in red.  The locations of the markers are: Tarfaya1 $22^\circ 22'10''$N $12^\circ 34'16''$W, Tarfaya2 $27^\circ28'55''$N $12^\circ33'25''$W, Tarfaya 3 $27^\circ23'50''$N $12^\circ33'31''$W, Mauritania1 $19^\circ12'20''$N $14^\circ23'37''$W, Mars1 $75^\circ1'0''$N $72^\circ8'0''$W, and Mars2 $73^\circ46'0''$N $70^\circ38'37''$W.  Images courtesy of Google Earth.}
    \label{fig: ThreeAreas}
\end{figure}

Each zone contained at least 1000 bedforms, including some complex dunes, which appear to be the intermediate stages of collisions, and bedforms for which no slip-face was visible: proto-dunes or dome dunes.  Imagery of the six zones are shown in figure \ref{fig: ThreeAreas}.  We chose to define all the zones as quadrilaterals however this choice was arbitrary and not required to maintain the assumption of periodic boundaries due to the homogeneity of large swarms \cite{elbelrhiti2008barchan}.

\subsection{Measuring dunes}

\begin{figure}
    \centering
    \includegraphics[width = \textwidth]{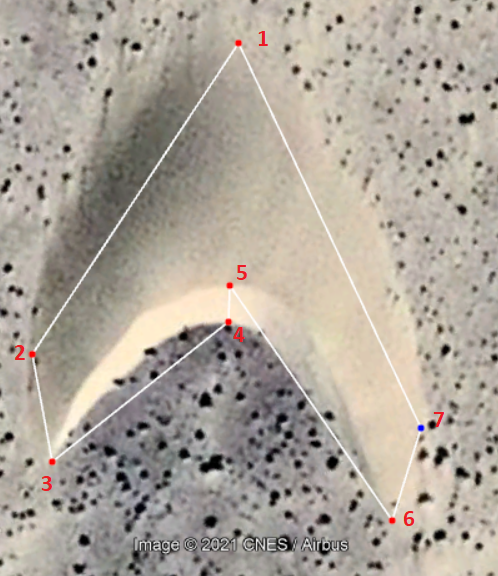}
    \caption{The seven points that were recorded on each barchan within the zones of study.  Image courtesy of Google Earth.  The dominant wind direction is from the from the top of the image to the bottom.}
    \label{fig: SevenPoints}
\end{figure}

We recorded seven locations on and around the body of every barchan-shaped bedform: the upwind toe, the leftmost edge, the tip of the left horn, the base of the slip-face, the brink, the tip of the right horn, and the rightmost edge (see figure \ref{fig: SevenPoints}).  From these points, it is possible to determine any of the linear morphological dimensions of the barchan.\\

\begin{figure}
    \centering
    \includegraphics[width = \textwidth]{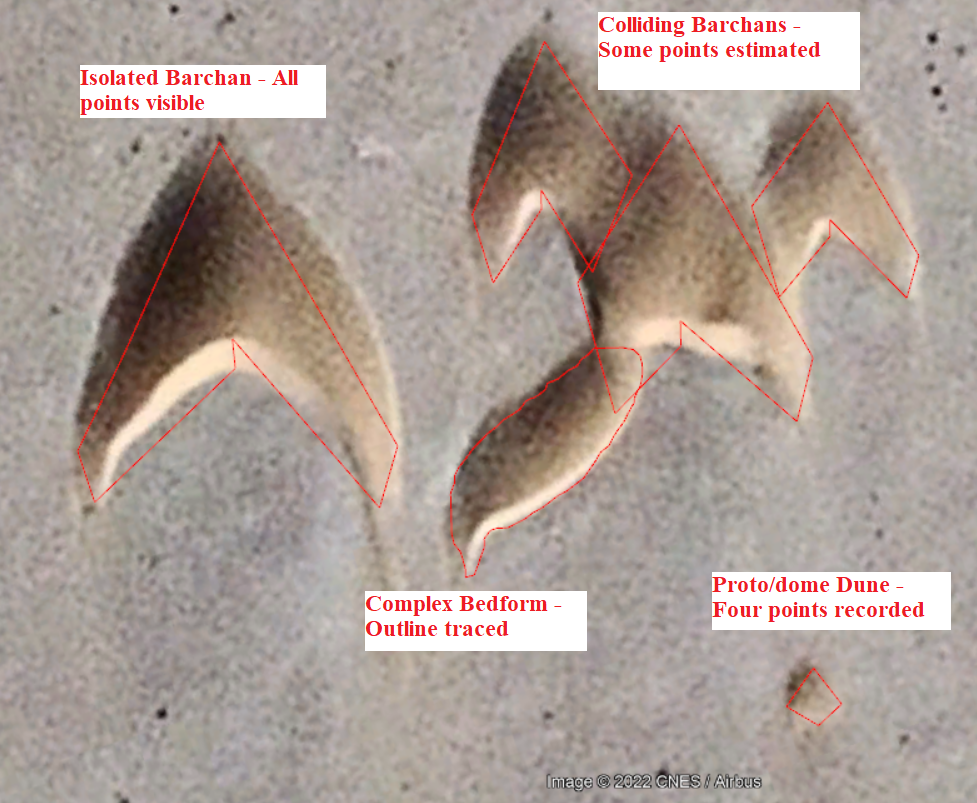}
    \caption{Examples of the measurement of: an isolated barchan (with all seven points visible), three colliding barchans (where some points were estimated), a complex bedform (for which the outline was traced), and a proto/dome-dune with only four points recorded. Image courtesy of Google Earth}
    \label{fig: Collision}
\end{figure}

Due to the high density of barchans in the zones of study, many of the barchans were not completely isolated bedforms but connected with another.  In these cases, it was not always possible to identify all seven points on the dunes in which case the locations were estimated as illustrated in \ref{fig: Collision}.  Such an attempt was made in all cases where there were two (or more) distinct crescentic slip-faces.  In cases where the leeward slope of a complex bedform exhibited some deformity but only one crescent was visible in the slip-face, the bedform was regarded as a single barchan.  Altogether,  connecting bedforms (including non-barchan overlapping bedforms) made up $40\%$, $36\%$, $34\%$, $7\%$, $4\%$, and $3\%$ of bedforms in Tarfaya zones 1-3, the Mauritania zone, and Mars zones 1 and 2 respectively.  In some cases, the bedforms were deformed to such an extent that, although a slip-face was visible, it was no longer crescentic.  Such dunes were not recorded as barchans and so no attempt was made to estimate the seven points; instead, we traced around the outline of the dune (see figure \ref{fig: Collision}).  These complex, highly deformed objects made up 2\%, 9\%, 5\%, 0\%, 2\%, and 2\% of the bedforms in Tarfaya 1-3, Mauritania and Mars 1 and 2 respectively.\\

In addition to complex bedforms resulting from dune-dune interactions, some bedforms did not have a visible slip-face.  When no slip-face was visible, but the shape was otherwise distinctly that of a barchan, it was assumed that the cause was poor image resolution and an attempt was made to estimate the location of all seven points.  Where there was no visible slip-face and the shape of the bedform did not appear barchan-like we recorded only four points: the upwind, downwind, leftmost, and rightmost extents of the bedform (see figure \ref{fig: Collision}).  We believe that these bedforms are likely to be predominately dome dunes or proto-dunes.  In Tarfaya 1-3, Mauritania, and Mars 1 and 2 these dome/proto dunes represented respectively 6\%, 9\%, 10\%, 2\%, 1\% and 1\% of the total bedform population. 

\subsection{Determining barchan volumes}

Various linear dune dimensions can be derived from the seven points recorded on each barchan, including: horn-to-horn width, total width, windward length, and total length.  However, to compare the outputs of the mean-field model to the observations of the real-world fields, it was necessary to estimate the volume of the bedforms (as this is the conserved quantity in the CAFE model) from the linear dimensions. Previous studies have shown how the volume of barchans scales as the cubic power of their height, length, and width \cite{elbelrhiti2008barchan, hersen2005flow, franklin2011subaqueous, zhang2010morphodynamics, duran2010continuous}.\\

We found that the most reliable linear dimension was the total length $l$, the distance between the upwind toe and foot of the slip-face (points 1 and 4 shown in figure \ref{fig: SevenPoints}).  The length was the easiest dimension to extract since it did not require correction for the orientation of the dune (as would be necessary for determining width).  We also found that there was a greater degree of subjectivity in locating exactly the position of the widest point, and sand streaming off of the horns sometimes made defining the tip of the horn difficult.  On the other hand, we found that, in most cases, the toe and slip-face were easily identified, making the length measurement more reliable than other dimensions.  To calculate the volume $v$ we used the relation $v = l^3/20$ which is a good description of the barchans in the Tarfaya Province \cite{elbelrhiti2008barchan}.

\subsection{Non-barchan bedforms}

\begin{figure}
    \centering
    \includegraphics[width = \textwidth]{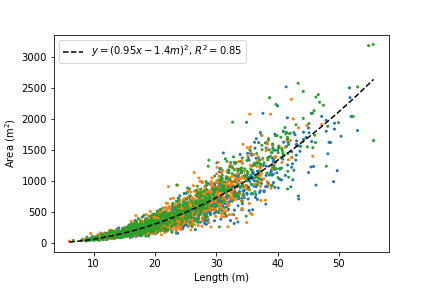}
    \caption{Observed basal areas of barchans in the three Tarfaya zones and their corresponding lengths.  The quadratic line of best fit and corresponding coefficient of determination are also shown.}
    \label{fig: AreaLengthScaling}
\end{figure}

Overall, barchans represented $\sim 91\%$ of the bedforms that we measured.  Although this represents a significant portion of the bedforms, the omission of non-barchan bedforms would still introduce a bias into our results.  Therefore, it was necessary to estimate the volumes of these other bedforms so that they could be included in determining the size distributions.  However, unlike well-defined barchans, there is no simple rule for calculating the volume of a complex bedforms.  To estimate the volumes, we instead assumed that the scaling of volume with basal area was the same for all bedforms.  We could then estimate the volume of a non-barchan bedform by finding the the size of a barchan with an equivalent basal area.  We found that the area $A_{b}$ of barchans was well modelled as $A_{b} = (c_1 l + c_2)^2$ with with: $c_1 = 0.95$ and $c_2 = -1.4m$ for Tarfaya, $c_1 = 0.88$ and $c_2 = 3.8m$ for Mauritania, and $c_1 = 0.65$ and $c_2 = 9.4m$ for Mars (with corresponding coefficients of determination  $R^2 = 0.85, \ 0.69$, and 0.69).  As an example, we show the fit for the Tarfaya barchans in figure \ref{fig: AreaLengthScaling}, the errors were symmetric about the line of best fit shown and displayed a symmetrical sub-Gaussian distribution.

\section{Results}\label{sec: Results}

We describe the observed size distributions, the methods by which we optimised model parameters, and the quality of the fits obtained using these optimal configurations. 

\subsection{Observed size distributions}

Including non-barchan bedforms, the population sizes for Tarfaya1, Tarfaya2, Tarfaya3, Mauritania, Mars1, and Mars2 were respectively 1015, 1357, 1008, 1009, 1001, 1002 making a total of 6392 measured bedforms.  Of these, the number of barchan bedforms in each of the six zones were respectively 927 ($91\%$), 1112 ($82\%$), 850 ($84\%$), 984 ($98\%$), 977 ($98\%$), and 975 ($97\%$) giving a total percentage of barchan bedforms of $91\%$.  Although our modelling focuses on the volume distributions, we first show the distributions of the lengths of the barchan bedforms for comparison with previous studies, before moving on to discuss the volume distributions when the other bedforms are included.

\subsubsection{Barchan length distributions}

In the four terrestrial swarms, the average length of barchans was approximately equal; Tarfaya zones 2 and 3 and the Mauritanian zone all had an average length of 23m while the barchans in Tarfaya1 were slightly larger with an average of 26m.  We observed that the Martian barchans were significantly larger than those on Earth with average lengths of 131m and 113m in Mars zones 1 and 2 respectively.  The large discrepancy between dune sizes on Earth and Mars is possibly caused by differences in grain-scale saltation mechanics in the two environments \cite{parteli2007minimal} as well as the external control of sediment availability.\\

 As average size of bedforms is likely controlled by external environmental factors and our study focuses on the internal dynamics of barchan-barchan interactions, we normalise data by the mean size of barchans in each zone allowing us to compare the distributions about the mean from all swarms.

\begin{figure}
    \centering
    \includegraphics[width = 0.48\textwidth]{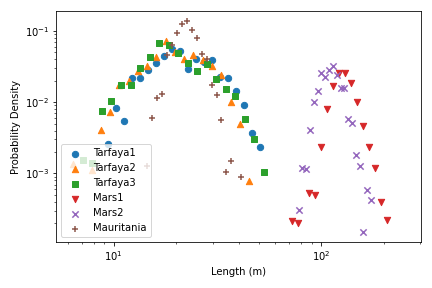}
    \includegraphics[width = 0.48\textwidth]{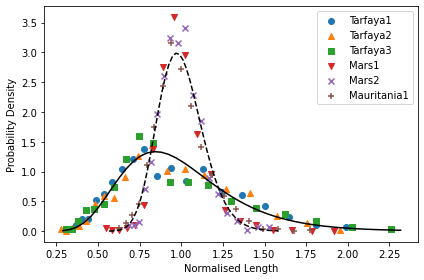}
    \caption{Left) Probability density functions are shown for the lengths of barchans in each of the six zones of study.  Right) The same distributions rescaled by the average length in each zone.  The lines show log-normal distributions estimated from the mean and variance of the observed data.}
    \label{fig: LengthDistributions}
\end{figure}

Figure \ref{fig: LengthDistributions} shows distributions of normalised length in each of the zones of study.  The distributions in the three Tarfaya zones are very similar which is unsurprising since these three zones constitute different areas within one extensive regional dune field. The similarity in these three distributions and in average size across the three Tarfaya zones suggests that our assumption of homogeneity in large fields is valid and is in agreement with previous findings \cite{elbelrhiti2008barchan, duran2011size}.  A similar argument could also be made for the two Martian fields, however, it is surprising that the Mauritania zone has a very similar distribution to the fields on Mars despite the large difference in the average size.  The striking similarity between the Mauritanian and Martian data suggests that the internal dune-dynamics in the two locales are similar despite the environmental conditions being very different.  These findings also potentially suggest that there may be a degree of universality of size distributions, with different classes of barchan fields (e.g. Tarfaya-type, Mars-Mauritania-type).\\

Although the distributions in the Tarfaya zones are rather different to those in Mauritania and Mars, all of the length distributions can be well-approximated by log-normal distributions as has been widely reported in previous studies \cite{elbelrhiti2008barchan, duran2009dune, duran2011size}. This can be seen in figure \ref{fig: LengthDistributions} where we show log-normal distributions fitted using the method of moments.

\subsubsection{Volume distributions}

\begin{figure}
    \centering
    \includegraphics[width = \textwidth]{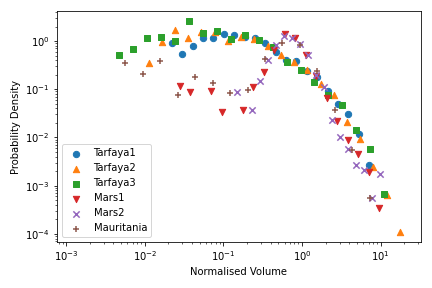}
    \caption{Probability densities of the estimated volume, normalised by the mean volume, of all bedforms (including non-barchan bedforms) in each of the six zones of study.}
    \label{fig: VolumeDistributions}
\end{figure}

The distributions of normalised volume for the six zones are shown in figure \ref{fig: VolumeDistributions}(note the log-log scale is necessary as the the spread of sizes is much greater than for the length distributions).  The similarity between the three Tarfaya zones and between the Martian and Mauritanian zones persists for the majority of the range of volumes, however slight differences appear on the smaller end of the distributions.  The similarities reflect the fact that the bedforms of intermediate size are likely to be true barchans or complex bedforms with similar sizes to the barchans (e.g. the complex bedform in figure \ref{fig: Collision}).  On the other hand, at the lower end of the volume range, the distributions are likely dominated by dome or proto-dunes which are more ephemeral and short-live due small size, thus exhibiting more variation between different regions within a large dune field such as between the different Tarfaya zones.  The relative abundance of small-volume bedforms in the Mauritanian zone compared to the Martian zones may be a reflection of the lower resolution of the Martian imagery we used.   Despite these differences at lower volumes, the data still demonstrate the striking property of two different classes of distribution.

\subsection{Free parameters and optimisation procedure}

As described in section \ref{sec: Rates}, interactions in the  {CAFE} model are governed by: the distributions of the stochastic variables $r_c$, $r_{f_1}$, $r_{f_2}$, and $r_e$; and the rate coefficients $\alpha_c$, $\alpha_a$, $\alpha_f$, and $\alpha_e$.  In all cases, $\alpha_a$ and $\alpha_f$ are not independent and therefore we have only three free rate coefficients.  This constraint on a rate coefficient is the only constraint that exists in the  {CAFE} model in general.  However, it is not possible to perform a generalised optimisation of the model since infinitely many free parameters would be required to describe all of the possible distributions for the stochastic variables.  To perform the optimisation, we chose to focus on two implementations of the model:
\begin{itemize}
    \item The empirical {CAFE} model where: $\alpha_c \neq 0$, $\alpha_a = 2 \alpha_c/(N - 1) + \alpha_f$, the distributions of $r_{f_1}$ and $r_e$ are determined from a known collision rule \cite{duran2009dune}, and the distributions of $r_c$ and $r_{f_2}$ are uniformly distributed in $[r_{c_{min}}, r_{c_{max}}]$ and $[r_{f_{2_{min}}}, r_{f_{2_{max}}}]$ respectively.
    \item The uniform exchange-only (UE) model where: $\alpha_c = \alpha_a = \alpha_f = 0$ and $r_e$ is distributed uniformly in the range $[r_{e_{min}}, r_{e_{max}}]$
\end{itemize}

The empirical CAFE model has seven parameters while the UE model has only two. The collision rule used to determine the distributions of $r_{f_1}$ and $r_e$ in the empirical CAFE model is well-established and has been used to model exchange interactions in barchan swarms in several previous studies \cite{duran2010continuous, duran2009dune, duran2011size}. A previous model by Worman et al. \cite{worman2013modeling} which included calving of barchans assumed that: a) only large dunes could calve, b) the small dunes formed by calving had a fixed size. While there is some anecdotal evidence that the size of a dune affects its ability to shed calving dunes \cite{elbelrhiti2008barchan, elbelrhiti2005field} quantitative empirical data is lacking \cite{worman2013modeling}. Furthermore, the imposition of a minimum size in \cite{worman2013modeling} led to a strong peak in their size distributions at this calving threshold, which is not observed in real-world distributions. We therefore do not impose size criteria in the CAFE model, allowing all dunes to calve.\\

The second assumption of the previous model \cite{worman2013modeling} was based upon linear instability analysis of surface waves on barchans \cite{elbelrhiti2005field}. Observed distributions of wavelengths of surface waves do show strong peaks \cite{elbelrhiti2005field} but these distributions are not consistent with observed sizes of proto/dome dunes \cite{elbelrhiti2005field} which have been observed to form due to calving \cite{bourke2010barchan, kinglibya1918}.  Furthermore, claims that all calved dunes ought to have a fixed size \cite{elbelrhiti2012initiation, worman2013modeling} are not consistent with continuum simulations of the calving process \cite{khosronejad2017genesis}. In the CAFE model, we restrict calving so that the width of the shed dune is less than 25\% of the source dune following calving. This upper limit is in line with the fixed sizes in the previous modelling \cite{worman2013modeling}. The resulting size distributions of calved dunes in the CAFE model are peaked but with some variation which is consistent with empirical evidence and continuum simulations \cite{elbelrhiti2005field, bourke2010barchan, kinglibya1918, khosronejad2017genesis}. The empirical CAFE model, therefore, represents a realistic approximation of calving and interactions in a real-world swarm. On the other hand, the simplicity of the UE model allows us to demonstrate how easily the observed distributions can be produced without recourse to relatively complex rules for interactions.

\begin{table}
\centering
\begin{tabular}{cccll}
\multicolumn{1}{l}{Model} & $\alpha_c$     & $\alpha_a$                                & $\alpha_f$            & $\alpha_e$ \\ \hline
 Empirical {CAFE}             &  $\geq0$        & $2\alpha_c/(N - 1) + \alpha_f$ & $\geq0$                  & $\geq 0$ \\    
 {UE} & 0 & 0 & 0 & 1
\end{tabular}
\caption{The rate coefficients in the empirical  {CAFE} and {UE} models.  Where a value is shown, the parameter is constrained to take that value.  An inequality denotes the constraint on a parameter which was otherwise free to be optimised.}
\label{tab: RateCoeffs}
\end{table}

\begin{table}
\centering
\begin{tabular}{cllll}
\multicolumn{1}{l}{Model} & $p_c(r_c)$                                 & $p_{f_1}(r_{f_1})$                                     & $p_{f_2}(r_{f_2})$                                    & $p_e(r_e)$                                 \\ \hline
 Empirical {CAFE}             & $[r_{c_{min}}, r_{c_{max}}]$ & Empirical                                     & $[r_{f_{2_{min}}}, r_{f_{2_{max}}}]$ & Empirical  \\
 {UE} & n/a & n/a & n/a & $[r_{e_{min}}, r_{e_{max}}]$
\end{tabular}
\caption{The distributions of the four stochastic variables in the empirical  {CAFE} and {UE} models.  We have used the shorthand $[x, y]$ to denote a uniform distribution in the range $[x,y]$.  In all such cases, $x$ and $y$ are parameters to be optimised, in all cases at least one constraint existed, namely $x, y \geq 0$.  For $r_{c_{min, max}}$ and additional constraint $r_{c_{min, max}} < 1/64$ was also imposed.  The empirical distributions were derived from a known collision rule.}
\label{tab: StochasticDists}
\end{table}

\subsubsection{Empirical collision rule}

In the empirical {CAFE} model we fix the distributions $p_{f_1}(r_{f_1})$ and $p_e(r_e)$ to a distribution we derived from an empirical interaction rule \cite{duran2010continuous, duran2009dune, duran2011size}.  This interaction rule was established by empirically fitting the volume ratio of outputs of barchan-barchan collisions modelled in continuum simulations to a function of their lateral offset and initial volume ratio \cite{duran2009dune, duran2010continuous}.  The details of the rule can be found in the appendix of \cite{duran2011size}.  The rule has previously been used in mean-field modelling for primarily exchange interactions \cite{duran2009dune}, however, we also took the ratio $r_{f_1}$ to be given by the same rule since many fragmentation interactions are similar to an exchange interaction where one of the outputs subsequently breaks apart.\\

The empirical collision rule is deterministic given the initial volume ratio and lateral offset \cite{duran2011size}.  To implement this rule in the  {CAFE} model we converted the deterministic expression into a probability distribution of outputs using the following procedure:

\begin{enumerate}
    \item Randomly select two dunes from our steady-state population and calculate their volume ratio $r_{in}$.
    \item Generate a random value for the lateral offset $\theta$ from a uniform distribution in the range $[0,1]$.
    \item Insert $r_{in}$ and $\theta$ into the deterministic collision rule \cite{duran2011size} to give the output volume ratio $r_{out}$.
    \item Record $r_{out}$ and repeat until $10^6$ values have been generated.
\end{enumerate}

\begin{figure}
    \centering
    \includegraphics[width = 0.49\textwidth]{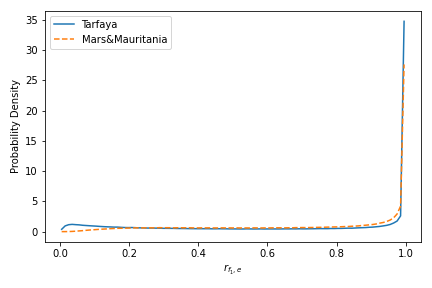}
    \includegraphics[width = 0.49\textwidth]{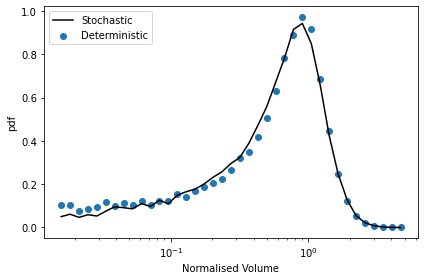}
    \caption{Left: The distributions used for $r_e$ and $r_{f_1}$ in the empirical  {CAFE} model for Tarfaya datasets (solid line) and the Mars and Mauritania datasets (dashed line).  Right: A comparison of the steady-state distribution produced using the deterministic collision rule \cite{duran2011size} and the stochastic collision rule we derived from the deterministic one.}
    \label{fig: StochVsDet}
\end{figure}

When running the empirical  {CAFE} model, each time we were required to generate a random value for $r_e$ or $r_{f_1}$ we simply randomly selected one of these $10^6$ deterministically calculated $r_{outs}$.  The resulting distributions for our datasets are shown in the left part of figure \ref{fig: StochVsDet}.\\

Before we proceeded with implementing this method using our real-world data, we first verified that the steady-state size distribution generated when using our stochastic interaction rule were the same as those generated when using the deterministic rule itself.  An example of this is shown in the right part of figure \ref{fig: StochVsDet}

\subsubsection{Optimisation}\label{sec: Optim}

Since we can solve for the steady-state distribution for a particular configuration of the  {CAFE} model, one method of optimising the parameters could be to use some measure of similarity between distributions, such as the Kolmogorov-Smirnov (KS) test statistic, as a cost function of an optimisation routine.  However, the time taken to solve for the steady-state, and the high dimensionality of the parameter space, mean we would only be able to explore a small area of the space using this technique.  Instead, we chose to make use of the fact that we have analytical expressions for all of the integer moments (equation \eqref{eq: Moments}).  The computational time taken to solve for the moments of the steady-state is, as such, much less than solving for the entire steady-state.  We therefore defined a cost function as the average percentage difference between the first 9 non-trivial moments of the observed and theoretical normalised volume distributions (since the volumes are normalised, the first moment is unity) i.e.

\begin{equation}\label{eq: CostFunction}
    \lambda = \frac{1}{9}\sum_{\ell = 2}^{10} \sqrt{\frac{(\mathbf{E}[v^\ell] - \langle v^\ell \rangle_s)^2}{\mathbf{E}[v^\ell]^2}}, 
\end{equation}

\noindent{where $\mathbf{E}[v^\ell]$ are the observed moments and $\langle v^\ell \rangle_s$ are the theoretical steady-state moments for a given set of parameters, calculated using equation \eqref{eq: Moments}.  This cost function minimises the average percentage error of the moments which, since the moments are rapidly increasing, means that the cost function prefers models where the lower moments are close to the data and puts less constraint on the higher moments.  This is important since the lower moments are more stable to the presence of large outliers.  We also found that implementing the generalised method of moments using weightings calculated from a heteroskedasticity-consistent covariance matrix did not converge to an optimum configuration within the permitted range of parameters (e.g. $r_{e_{min}} \geq 0$ etc.) }\\

The optimisation of the empirical {CAFE} model then consisted of finding the global minimum of our cost function.  We performed this minimisation using the SciPy \cite{2020SciPy-NMeth} implementation of dual annealing, a technique which couples generalised annealing with a local search \cite{tsallis1988possible, tsallis1996generalized, xiang1997generalized, xiang2000efficiency, xiang2013generalized}.\\

Optimisation of the UE model was considerably easier than the empirical {CAFE} model for two reasons: 1) the UE model has only 2 free parameters, 2) our choice of using a uniform distribution for $r_e$ means that the left-hand side of equation \ref{eq: KL} can be written as an analytical expressions of $r_{e_{min}}$ and $r_{e_{max}}$.  The second point is particularly important as we have previously shown in \cite{robson2021combined} that the steady-state of exchange-only systems is well-approximated by a gamma distribution with parameters that can be directly calculated from $K_2$.  The optimisation of the UE model parameters was then a two-step process:

\begin{enumerate}
    \item Estimate $K_2$ from the moments of the real-world distribution as
    \begin{equation}
    K_2 = \frac{\mathbf{E}[v^2]}{2\left(\mathbf{E}[v]^2 + \mathbf{E}[v^2]\right)}
    \end{equation}
    \item Iteratively solve
    \begin{equation}
        r_{{max}} - r_{{min}} = \frac{\log\left(\frac{1 + r_{{max}}}{1 + r_{{min}}}\right)}{\frac{1}{2} + \frac{1}{(1 + r_{{max}})(1 + r_{{min}})} - K_2},
    \end{equation}
    where we have used the shorthand $r_{e_{min, max}} = r_{min, max}$.
\end{enumerate}

We found that the values of $r_{e_{min, max}}$ converged to a precision of 0.001 after $\sim 40$ steps.  It is worth noting that, there are often multiple possible solutions for the two parameters, however, we are interested primarily in reproducing the size distributions and so we only require that the algorithm converges to a solution.\\

\subsection{Optimum parameters}

We ran the optimisation algorithms for each of the zones individually and for combinations of the Tarfaya data and the Mars and Mauritania data.  We will focus on the results of these combined optimisations since the larger population sizes in combined datasets reduced the impacts of outliers on our results.  Furthermore, the slight differences between otherwise similar distributions may give a more accurate description of the overall steady-state distribution since each individual zone represents only a snapshot and therefore may display features which are the result of a fluctuation.  We focus, therefore, on the two different classes of distributions we have observed namely: 1) the combination of the three Tarfaya zones 2) the combination of the Mars and Mauritania zones. Although the effects of outliers is reduced when the populations are combined, we find that extreme values in the Mars and Mauritania datasets have a large influence on the optimum model parameters and that a much better fit is achieved when the bottom and top 1 percentiles of the distribution are ignored when finding the optimum parameters.\\

 To aid comparison between the rate coefficients for calving and the three interaction processes, we define \textit{event probabilities} $\rho_i$ with $i = c, \ a, \ f, \ \text{or} \ e$

\begin{align}
    \rho_c &= \frac{\alpha_c N}{\alpha_c N + (\alpha_a + \alpha_f + \alpha_e) N (N-1)/2},\\
    \rho_{a, f, e} &= \frac{\alpha_{a,f,e} N (N-1)}{2\alpha_c N + (\alpha_a + \alpha_f + \alpha_e) N (N-1)}.
\end{align}

\noindent{These event probabilities, $\rho_{c,a,f,e}$, are the probabilities that the next event (calving or collision) to occur in the system will be calving, aggregation, fragmentation, or exchange accounting for the different dependence of the rate coefficients on population size.  The event probabilities can also be calculated from the channel probabilities define in section \ref{sec: Model} with}

\begin{align}
    \rho_c &= p_1 + p_2 =  2 p_{1,2},\\
    \rho_a &= 2 p_3,\\
    \rho_f &= 2/3 (p4 + p5 + p6) =  2 p_{4,5,6},\\
    \rho_e &= p_7 + p_8 = 2 p_{7, 8}.
\end{align}

\noindent{Note that since these four events are the only processes in our model $\rho_c + \rho_a + \rho_f + \rho_e = 1$, and that, in any model, $\rho_c + \rho_f = \rho_a$ since a steady-state is only possible if the processes leading to an increase in population size are in equilibrium with those that decrease population size.  For the empirical CAFE model, the optimum event probabilities and stochastic distribution parameters are shown in table \ref{tab: ECAFEParams}.  For the UE model, by definition, $\rho_e = 1$ while all other event probabilities are zero.}\\

\begin{figure}
    \centering
    \includegraphics[width = 0.49\textwidth]{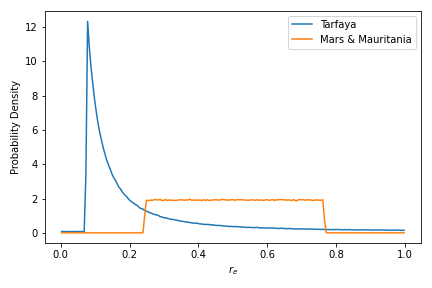}
    \includegraphics[width = 0.49\textwidth]{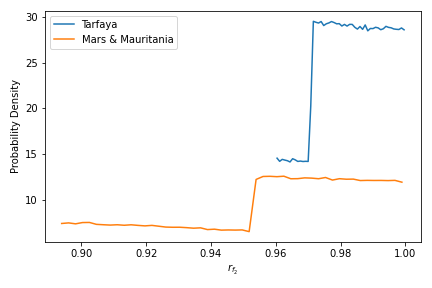}
    \caption{The optimum distributions of the stochastic variables $r_e$ (in the UE model) and $r_{f_2}$ in the empirical CAFE model for the Tarfaya dataset and the truncated Mars-Mauritania dataset.  Left: The optimum distributions of $r_e$ in the UE model.  Right: the optimum distributions of $r_{f_2}$ in the empirical CAFE model.}
    \label{fig: OptStochs}
\end{figure}

As described in table \ref{tab: StochasticDists}, the UE model only requires the two parameters that define the range of the uniform distribution of $r_e$.  For Tarfaya, we find the optimum range to be $r_e \in [0.00, 13.5]$ while for the truncated Mars and Mauritania dataset, the optimum distribution is $r_e \in [0.244, 0.766]$.  Note that, in the UE model we do not impose an upper limit on $r_{e_{min, max}}$, similarly we did not add an upper constraint on $r_{f_{2_{min,max}}}$ in the empirical {CAFE} model.  Since these values are ratios, they would typically be thought to exist in the range $[0, 1]$.  By allowing the parameters to take values greater than unity, we can effectively introduce a peak into their distribution without having to move away from the uniform distribution (which is the easiest to deal with analytically).  This is because assigning a ratio of greater than one is equivalent to swapping the numerator and denominator.  Therefore, values of $r_e > 1$ in the UE model and $r_{f_2} > 1$ in the empirical CAFE model can be converted to $1/r_e < 1$ and $1/r_{f_2} < 1$ without altering the outcome of the interaction.  In figure \ref{fig: OptStochs} we show the optimum distributions for the stochastic variables in our models with the transformation we have just described applied whenever a value is greater than unity.  Figure \ref{fig: OptStochs} does not include the distributions of $r_e$ and $r_{f_1}$ in the empirical CAFE model as those have already been shown in figure \ref{fig: StochVsDet}

\begin{table}[ht]
\begin{tabular}{lccccccll}

Zone & $\rho_c$(\%) & $\rho_a$(\%) & $\rho_f$(\%) & $\rho_e$(\%) & $r_{c_{min}}$ & $r_{c_{max}}$ & $r_{f_{2_{min}}}$ & $r_{f_{2_{max}}}$ \\ \hline
Tarfaya & $0.121$ & $48.2$ & $48.1$ & $3.61$ & $3.44 \times 10^{-4}$ & $4.78 \times 10^{-4}$ & 0.960 & 1.03 \\
M.\&Mau. & $1.43 \times 10^{-3}$ & $6.53$ & $6.53$ & $86.9$ & $0.0147$ & $0.0155$ & 0.953 & 1.12
\end{tabular}
\caption{The optimum event probabilities and variable parameters of the empirical {CAFE} model for the combined Tarfaya dataset and the truncated combined Mars and Mauritania (M.\&Mau.) dataset.  The truncation of the Mars and Mauritania data involved removing the bottom and top 1 percentiles.}
\label{tab: ECAFEParams}
\end{table}

\subsection{Steady-state distributions}

\begin{figure}
    \centering
    \includegraphics[width = 0.49\textwidth]{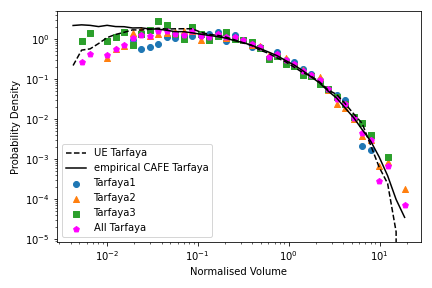}
    \includegraphics[width = 0.49\textwidth]{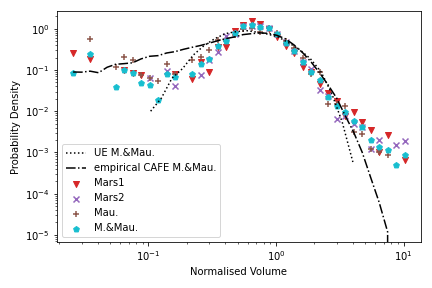}
    \caption{The normalised steady-state volume distributions corresponding to the optimised parameters for the empirical CAFE and UE models.  Left: The models were optimised to fit the the combined data set (pentagons).  Right: The models were fit to the combined dataset with the top and bottom 1 percentile removed to yield a better fit.}
    \label{fig: SteadyStateFits}
\end{figure}

The optimum model parameters were inserted into a population dynamics algorithm \cite{robson2021combined, mezard2001bethe, agliari2013immune} to find the corresponding steady-state distributions.  In figure \ref{fig: SteadyStateFits} we show the volume distributions of steady-states for the optimised parameters calculated for the combined Tarfaya dataset and the combined Mars and Mauritania dataset, alongside the observed distributions in the individual zones. \\

To quantitatively evaluate the strength of each fit we calculated two-sample KS statistics from the raw datasets and the steady-state outputs of the three models.  The values of the test-statistic for the combined Tarfaya data and the combined Mars and Mauritania data compared to both the empirical CAFE and UE models are shown in table \ref{tab: KSStats}.  (note that, we compare the steady-state of the models to the full Mars-Mauritania data including the top and bottom 1 percentiles).  In all cases, the KS test statistics are large enough to reject the null hypothesis that the model steady-state and the observed data are from the same distribution.

\begin{table}
\centering
\begin{tabular}{lcc}
Zone & empirical CAFE & UE \\ \hline
Tarfaya & 0.0524 & 0.0526 \\
Mars\& Mauritania & 0.103 & 0.0937
\end{tabular}
\caption{Two-sample Kolmogorov-Smirnov (KS) statistics calculated between the observed distributions and the optimised model steady-states for the two models and the two combined datasets.}
\label{tab: KSStats}
\end{table}

\section{Discussion of results}\label{sec: Discuss}

The resemblance of Martian and Mauritanian datasets is surprising.  Given the large differences between the environments of Mars and Mauritania their close fit strongly suggests
that it is indeed the internal processes of barchan-barchan interactions that govern the shape of the size distributions, with external controls determining the average size.  It also appears that there are attractor states for the steady-state size distributions which suggests that the processes are not free but conform to set configurations.\\

The empirical CAFE model represents a more realistic approximation of calving and interactions within barchan swarms.  We find that the Tarfaya distribution can be fairly accurately reproduced using the empirical CAFE model although the differences between the model steady-state and the observed distribution are statistically significant.  The replication of the Mars-Mauritania distribution is less convincing, although the fit of the data is much more accurate away from the tails of the observed distribution.  The fits produced using the simpler but less realistic UE model are marginally better than those produced by the empirical CAFE.  However, again, there is a statistically significant difference between the steady-state distributions and the real-world size distributions.\\

The optimum distributions of the stochastic variable $r_{f_2}$ in the empirical CAFE model are very similar for both the Tarfaya data and the Mars-Mauritania data with uniform distributions favouring $r_{f_2} \sim 1$ in both cases.  That the same optimum configuration is observed in the different regions on Earth as well as Mars suggests that this may be a universal description of the nature of fragmentation interactions between barchans.  Recalling that, in the empirical CAFE model, the distribution of $r_{f_1}$ also favours $r_{f_1} \sim 1$ (see figure \ref{fig: StochVsDet}), our findings suggest that fragmentation interactions typically produce two approximately equal sized dunes and a third dune with a volume equal to the sum of the other two, i.e. volumes in the ratio 1:1:2.  Compared to previously reported types of fragmentation, this is perhaps most similar to the ``budding'' described in \cite{duran2005breeding}.\\

For calving, contrast in the distributions of $r_c$ between the two cases is perhaps due to the parameters being poorly constrained as, in both cases, calving is found to be very rare.  In the Mars-Mauritania case, the calving parameter range optimises close to the imposed limit of 1/64, representing a calving dune with a width a quarter of the remaining dune, whereas in the Tarfaya case the optimised calving rule represents a calved dune that is only $\sim 7\%$ the width of the remaining dune.\\

While interaction rules in the empirical CAFE model are consistent for the two swarm types (except for the very rare calving events),  the optimised event probabilities are very different for the Tarfaya data and the Mars-Mauritania data.  In Tarfaya we find that a good fit to the data can be achieved when nearly all ($>96\%$) events are either aggregation or fragmentation collisions, with aggregation only marginally preferred.  On the other hand, in Mars and Mauritania we find that exchange interactions are dominant ($\sim 87\%)$ while aggregation and fragmentation make up the remaining $\sim 13\%$.  The different probabilities are consistent with the much greater relative spread of volumes in the Tarfaya zones than those in Mars and Mauritania.  Aggregation promotes a small number of large dunes while fragmentation promotes a large number of smaller dunes, and thus, if both processes occur frequently. a large overall spread of dune sizes can be maintained.  On the other hand, the empirical rule for exchange interactions is strongly peaked close to unity meaning that exchange often leads to similarly sized dunes which favours a strongly peaked distribution such as seen for intermediate volumes in the Mars and Mauritania data and as reported in mean-field modelling using the deterministic rule on which our empirical rule was based\cite{duran2009dune}.  This also explains why the fit of volumes close to the peak of the Mars and Mauritania distributions is much better than in the tails.  In the Tafaya zones, the wide spread of volumes promoted by aggregation-fragmentation leads to significant variation in the dune migration rates with the large products of aggregation moving much more slowly than the small barchans formed during fragmentation \cite{elbelrhiti2008barchan, hersen2005flow}.  The disparities in migration rates yields a much greater rate of collisions as smaller dunes catch up with larger ones.  This is commensurate with our findings that a much greater proportion of bedforms are in a connected stage, i.e. in the process of colliding, in Tarfaya (34-40\%) compared to the zones on Mars (3-4\%) and in Mauritania (7\%).\\\

The optimal configuration of parameters in the UE model similarly reflect the differences in the range of volumes in each of the two cases.  For the Tarfaya data the optimal distribution for $r_e$ is effectively peaked at $r_e \sim 0.074$ since we found an optimum $r_{e_{max}} = 13.5 = 1/0.074$.  With this rule, exchange interactions generally produce one dune that is only around $7\%$ of the volume of the other thus maintaining the dune size variation.  For the Mars-Mauritania dataset, however, the ratio of output dunes is ranges between 25\% and 75\% reflecting the fact that the peak in the Mars-Mauritania data is much stronger than in the Tarfaya data.  In the UE model, the differences in the size-distributions can only be achieved by using divergent exchange rules (conflicting distributions of the stochastic variable $r_e$) for different swarms.\\

In summary, for the Tarfaya swarms we find wide size distributions that are maintained by predominantly aggregation-fragmentation dune interactions yielding a wide range of sizes, whereas in the Mars and Mauritania swarms we find peaked distributions maintained by predominantly exchange interactions yielding a narrow range of sizes. The empirical CAFE model appears the more appropriate descriptor for the distributions, as it captures physically motivated rules for known interactions that are, crucially, universal for both types of barchan swarms on both planets.\\

The origins of the differences in size distributions of the two types of barchan swarm cannot be directly determined from our model although there are some possibilities. The population dynamics evolution of the two types represent attractor states that reflect the proportion of aggregation-fragmentation interactions versus exchange interactions, independent of initial or boundary conditions. That is: starting with any size distribution for a nascent swarm or ingressing dunes the aggregation-fragmentation processes evolve a swarm to a specific wide size distribution that is subsequently maintained, just like the exchange process evolves a swarm to a specific peaked distribution that is then maintained. On this basis we can dismiss initial or boundary conditions as controls on the shape of the size distribution, although they are assumed to be responsible for the overall scale (average size) of the dunes in the swarm (the amount of sediment available to ‘build’ the dunes).\\

The predominance of aggregation-fragmentation processes in the Tarfaya swarms as opposed to the majority exchange interactions in Mars-Mauritania swarms may instead hint at key differences in transfer of mass between dunes via interdune sand flux. Sand flux is not included in the models we analysed, and it is possible that aggregation-fragmentation partly compensates for or reflects this feature: the process of very small dunes generated during fragmentation subsequently aggregating with other dunes may be interpreted as a sand flux transfer in the small size limit (recall that the models do not impose a minimum dune size threshold). While the visual evidence of many overlapping dunes in the Tarfaya suggests that aggregation and fragmentation are genuine dune interactions there, the differences in event probabilities between Tarfaya and Mars-Mauritania may also be an indicator that interdune sand flux is more important in the former than in the latter. We suggest this hypothesis for future investigation.\\

Spontaneous calving is also a process yielding very small dunes that could represent interdune sand flux, but its modelled event probability is very low in both types of swarms (though interestingly, nearly 100 times more likely at Tarfaya than in Mars-Mauritania swarms). Calving is a less efficient process for mimicking interdune sand flux than fragmentation, however, as the latter produces two small dunes as opposed to just one from calving. Our understanding of the dune interaction types and how they affect the size distribution does make clear that calving will mainly affect the tails of the distribution (the very small and very large dunes) and does not seem to play a role in size regulation, or the peakedness of the distribution, contrary to what previous studies have suggested \cite{worman2013modeling, elbelrhiti2008barchan}.  If calving does play a role in governing the sizes of barchans in swarms, our results suggest that it may be in the tails of the distribution where there is the greatest discrepancy between our model steady-states and the observed distributions.\\

Wind regime may play an important role in affecting the internal processes within a barchan swarm, however there is currently very little understanding about how the two are related. Some works have suggested that increased variability in the wind regime may lead to greater rates of calving \cite{elbelrhiti2008barchan, elbelrhiti2005field} but do not provide quantitative predictions. The influence of wind regime on interactions is even less well understood. There is some evidence from water tank experiments that the shear velocity may impact interactions \cite{assis2020comprehensive} however the problem has not been understood quantitatively and there is no empirical evidence of any impacts of wind on interactions in aeolian settings. The optimal parameters we observed may contain some information about the wind regimes affecting the swarms, but without a better understanding of the link between processes and wind regime any attempt to analyse along these lines would be purely speculative.\\

Although the CAFE model provides insight into the internal processes and interactions in barchan swarms in particular with the finding of a universal rule for the outputs of fragmentation interactions, there remain a number of open questions relating to the relevance of sand-flux and the role of calving in the tails of the distribution. Furthermore, there are other observed properties of barchan swarms, most notably the emergence of homogeneity and horn-to-horn alignment of nearby barchans \cite{elbelrhiti2008barchan}, which cannot be studied using the CAFE model, but which are influenced by the same processes. Future endeavours should therefore focus on understanding barchan swarms using spatially resolved models using the other observed phenomena to provide additional constraints on the interaction rules and avoiding some of the underlying simplifying assumptions of mean-field models.  The additional constraints may be particularly useful in identifying the role played by wind regime which, as discussed above, is not easily extracted from size distributions alone as the effects of wind regime on barchan-barchan interaction have not been rigorously analysed at this point.

\section{Conclusion}\label{sec: Conclusion}

Comparing the observed size distributions to a mean-field model (the CAFE model) featuring calving, aggregation, fragmentation, and exchange, we studied aeolian barchan swarms in six different study areas: two in the northern circumpolar region of Mars, three in Tarfaya, Morocco, and one in Mauritania.\\

We estimated the volumes of the dunes by using a scaling law for barchan length and basal area which we derived.  We find a striking similarity between the distributions in the Mars and Mauritania zones despite the significant differences in the planetary environments.  The three Tarfaya zones display a distribution that is distinct from the Mars-Mauritania distribution which suggests that internal dune dynamics produce attractor states i.e. Tarfaya-type and Mars-Mauritania-type.\\

A particular model implementation, the empirical CAFE model, is the most physically relevant and demonstrates that universal interaction rules can produce both the Tarfaya and Mars-Mauritania distributions.  Specifically, we find that exchange interactions most commonly produce two dunes of roughly equal size while fragmentation interactions lead to three dunes with volumes in the ratio 1:1:2.\\

The empirical CAFE model explains the differences between the Tarfaya and Mars-Mauritania distributions as resulting from differences in the relative frequency of the dune processes.  We find that the Mars-Mauritania distribution is maintained by a dominance of exchange interactions while the Tarfaya distribution results from aggregation-fragmentation dynamics.\\

We hypothesise that a greater level of interdune sand-flux in Tarfaya may explain some of the increased importance we find for aggregation-fragmentation dynamics.  However,  greater rates of genuine aggregation-fragmentation interactions would lead to an increased rate of collisions which is consistent with our observation that more bedforms are in the process of colliding in the Tarfaya zones than in Mauritania and on Mars.\\

We find that spontaneous calving is rare and does not play an important role in size regulation in either type of distribution.  Calving is likely only relevant for the tails of bedform size distributions.

\section*{Acknowledgments}

DTR is supported by the EPSRC Centre for Doctoral Training in Cross-Disciplinary Approaches to Non-Equilibrium Systems (CANES EP/L015854/1).  We would also like to thank the anonymous reviewers for their constructive feedback during the peer review process.

\newpage

\bibliographystyle{unsrt}
\bibliography{refs}
\end{document}